# Bandgap-Dependent Doping of Semiconducting Carbon Nanotube Networks by Proton-Coupled Electron Transfer for Stable Thermoelectrics

*Angus Hawkey, Xabier Rodríguez-Martínez\*, Sebastian Lindenthal, Moritz C. F. Jansen, Reverant Crispin, Jana Zaumseil\**

Angus Hawkey, Xabier Rodríguez-Martínez, Sebastian Lindenthal, Moritz C. F. Jansen, Jana Zaumseil
Institute for Physical Chemistry, Heidelberg University, 69120 Heidelberg, Germany

Xabier Rodríguez-Martínez

Universidade da Coruña, Centro de Investigación en Tecnoloxías Navais e Industriais (CITENI), Ferrol, 15403, Spain

Reverant Crispin

Laboratory of Organic Electronics, Department of Science and Technology, Linköping University, Sweden

E-mail: zaumseil@uni-heidelberg.de; xabier.rodriguez@udc.es

**Abstract**

Networks of semiconducting single-walled carbon nanotubes (SWNTs) are a promising material for thermoelectric energy harvesting due to their mechanical flexibility, solution processability, high Seebeck coefficients and high electrical conductivities after chemical p- or n-doping. Here, we demonstrate that proton-coupled electron transfer (PCET) with benzoquinone (BQ) as the oxidant and lithium bis(trifluoromethylsulfonyl)imide (Li[TFSI]) for electrolyte counterions is a promising method for p-doping of polymer-sorted semiconducting SWNT networks. The achieved doping levels, as determined from absorption bleaching, depend directly on both the pH of the aqueous doping solutions and the bandgap (i.e., diameter) of the nanotubes within the network. Fast screening of different nanotube networks under various doping conditions was enabled by a high-throughput setup for thermoelectric measurements of five samples in parallel. For small-bandgap SWNTs, PCET-doping is sufficient to reach the maximum thermoelectric power factors, which are equal to those obtained by conventional methods. In contrast to other doping methods, the electrical conductivity of PCET-doped SWNTs remains stable over at least 5 days in air. These results confirm PCET to be a suitable approach for more environmentally friendly and stable doping of semiconducting SWNTs as promising thermoelectric materials.

# 1. Introduction

Dense networks of semiconducting single-walled carbon nanotubes (SWNTs) are a promising thermoelectric material with competitive thermoelectric power factors (PFs) of up to 920 µW m$^{-1}$ K$^{-2}$ when chemically doped.[1-3] Due to their mechanical flexibility and solution-processability, they are also a potential alternative to the widely investigated polymer semiconductors for flexible and carbon-based thermoelectric generators.[4-5] The quasi-one-dimensional structure of SWNTs can be represented as a graphene sheet rolled into a seamless tube with diameters of 0.5 to 2 nm and lengths of hundreds of nanometers to few micrometers. The electronic (metallic or semiconducting) and optical (near-infrared absorption and emission) properties of SWNTs depend directly on the tube diameter and roll-up angle, which are described by a pair of chiral indices (n,m).[6] Importantly, the bandgap of semiconducting SWNTs is inversely proportional to their diameter and their intrinsic carrier mobilities scale with the square of the diameter.[7]

As-synthesized SWNTs contain a mixture of semiconducting and metallic species with different diameters, but purely semiconducting SWNTs are usually desired to maximize the thermoelectric performance of networks.[2, 8] One approach to separate semiconducting from metallic nanotubes is their selective dispersion with conjugated polymers, usually polyfluorenes, in organic solvents.[9-10] This polymer-sorting is a highly reproducible and scalable method that achieves >99.99% semiconducting purity.[11-12] Depending on the SWNT raw material and conjugated polymer, nanotubes of a single (n,m) type (monochiral) or distributions of semiconducting SWNTs within a certain diameter and hence bandgap range can be obtained. These polymer-wrapped SWNTs are stable in dispersion and can be processed into thin films by printing,[13] spin-coating,[14] or filtration[15] for applications such as electrochromic filters,[16] electronic circuits, [17] and thermoelectric generators.[2, 18]

Just like polymer semiconductors, SWNT networks must be p- or n-doped to create efficient thermoelectric devices. However, with increasing carrier density the Seebeck coefficient ($S$) decreases and at a certain conductivity ($\sigma$) a peak of the PF (with PF = $S^2\sigma$) is reached.[19] Many of the chemical dopants and doping strategies developed for organic and polymeric semiconductors[20-21] have also been applied to carbon nanotubes. This includes the immersion of SWNT films in solutions of oxidants (for p-type doping) such as 2,3,5,6-tetrafluoro-7,7,8,8-tetracyanoquinodimethane ($F_4TCNQ$),[22] triethyloxonium hexachloroantimonate (OA),[23] AgTFSI,[24] $FeCl_3$,[25] or $AuCl_3$[26] or reducing agents (for n-doping) such as crown-ether complexes of alkali metals[27] or benzimidazole derivatives.[28] The largest PF values for nanotube networks so far were observed for p-doping of small-bandgap SWNTs with functionalized icosahedral dodecaborane clusters, whose large counter anions improved charge transport in the doped nanotubes.[3] Importantly, while networks of SWNTs are fairly porous they are completely insoluble and do not swell or change their morphology during the application of doping solutions, which is another advantage compared to many organic semiconductors.

Important features of chemical doping of both SWNT networks and organic semiconductors are control of the precise doping level (ideally at the maximum PF) and doping stability. Due to the presence of the redox-active counterion of the dopant, the charge-transfer is often reversible and not permanent. Furthermore, due to the often required strong oxidizing potential of the p-dopant, side reactions may occur. Some strong dopants such as $FeCl_3$ can generate unstable radical species,[29] resulting in the degradation of the semiconductor. Several strategies were developed in the past years to mitigate these problems for organic semiconductors. Ion-exchange doping was introduced to replace the redox active counterion with a redox-inactive electrolyte anion or cation, which is added in excess to further drive the equilibrium toward charge transfer from the dopant to the semiconductor.[30-32] It also enables the investigation of the possible impact of different counterions on charge transport and Seebeck coefficients.[33] Ion-exchange doping was also successfully applied to polymer-sorted SWNTs.[15] While ion-exchange doping can replace unstable counterions with stable closed-shell electrolyte anions, some residual dopants still remain and over time degrade the thermoelectric performance.[15, 29]

Thus, recent research has focused on enhancing control over doping levels and improving the doping stability through new doping methods which avoid the use of harsh and unstable dopants. Jin *et al.* introduced photocatalytic doping, which utilizes air-stable photocatalysts to oxidize (p-dope) or reduce (n-dope) semiconducting polymers.[34] This method enables precise control over the doping level by controlling the light-irradiation dose. The photocatalyst is regenerated through a reaction with weak dopants, such as oxygen, resulting in stable conductivity for over 30 days. Ishii *et al.* demonstrated p-doping of semiconducting polymers via proton-coupled electron transfer (PCET) with aqueous solutions of quinones at different pH values.[35] In this approach, a quinone (e.g., benzoquinone, BQ) accepts two electrons from the polymer and two protons ($H^+$) from the buffered solution to form a neutral hydroquinone (HQ). The hydroquinone cannot serve as a counterion for the p-doped polymer and hence the presence of excess electrolyte anions (e.g., bis(trifluoromethylsulfonyl)imide, TFSI) provides charge compensation and drives the electron transfer. As described by the Nernst equation, the redox potential of the quinone increases as the pH is lowered, thus allowing for precise tuning of the doping level simply by adjusting the pH of the doping solution.[35] The lack of unstable counterions, precise tunability of the doping levels and use of aqueous solutions makes PCET a highly promising method for stable p-doping of thermoelectric materials.

Here, we apply PCET-doping with BQ to dense networks of polymer-sorted semiconducting SWNTs. We investigate the dependence of the doping efficiency on the bandgaps and thus oxidation potentials of the SWNTs and the pH values of the doping solutions. All doped networks are characterized with regard to their optical properties (absorption bleaching) and thermoelectric properties (conductivity, Seebeck coefficient, power factor). While PCET-doping emerges to be inadequate for large-bandgap nanotubes, for small-bandgap nanotubes it achieves the same electrical conductivities and power factors as ion-exchange doping with AuCl$_3$/TFSI. Importantly, PCET-doped SWNT networks also show excellent long-term stability.

## 2. Results and Discussion

To investigate the applicability and efficiency of PCET-doping for dense networks of SWNTs with different bandgaps, we selectively dispersed large amounts of semiconducting SWNTs from different sources with conjugated polymers in toluene by shear-force mixing or bath sonication as previously described.[9, 11, 36] To isolate small-diameter (6,5) and (7,5) SWNTs (diameter 0.76 and 0.83 nm, bandgap 1.27 eV and 1.21 eV, respectively), CoMoCAT SWNTs were dispersed and sorted using the established polyfluorene polymers PFO-BPy and PFO (for molecular structures see Figure 1a, respectively. For a mix of five different semiconducting SWNTs (diameter 0.83 ‑ 1.1 nm, bandgap 0.94 ‑ 1.21 eV), HiPco SWNTs were dispersed with PFO in toluene. Lastly, a broad mix of large-diameter semiconducting SWNTs (diameter 1.17 – 1.55 nm, bandgap 0.70 – 0.88 eV) were obtained from plasma torch (PT) SWNTs by dispersion with PFO-BPy (for complete absorption spectra of the dispersions see Figure S1). These dispersions were used to form dense films by vacuum filtration onto mixed cellulose ester (MCE) membranes. The thickness of the films was controlled by the volume and concentration of the filtered nanotube dispersions. Excess wrapping polymer was removed by washing in hot toluene, leaving only a minimal amount of polymer wrapped around the SWNTs. The residual polymer does not contribute to charge transport due to its large bandgap compared to the nanotubes.[37] The filtered SWNT films were cut to the desired dimensions and transferred from the MCE membranes onto glass substrates with up to 5 pairs of pre-deposited gold electrodes as described previously.[15] A photograph of such dense nanotube films on the same substrate is presented in Figure 1a. The different colors originate mainly from their second absorption peak ($E_{22}$) in the visible range (see Figure S1). Figure 1b shows the absorption spectra of the corresponding dispersions in the near-infrared region, where the main and lowest energy transitions ($E_{11}$) occur. These correlate roughly with the bandgap of the nanotubes. An atomic force microscopy (AFM) image of a typical transferred PT-SWNT film (see Figure 1c) confirms the density of these networks with only very little residual polymer.

As visualized in Figure 1d, the different nanotube networks represent different bandgaps from fairly large for the (6,5) and (7,5) nanotubes (> 1.2 eV) to smaller bandgaps (< 1 eV) for the semiconducting PT nanotubes. These differences can be employed to probe the doping efficiency of PCET with BQ at different pH values. The general principle of PCET doping is outlined again in Figure 1e. The BQ is reduced to HQ by a two-electron two-proton transfer. The electrons are provided by the SWNT network, which is oxidized (i.e., p-doped). As described previously by Ishii et al.,[35] the equilibrium of the reaction is driven by the excess concentration of the counter anions of the electrolyte (here TFSI) and the concentration of $H^+$ (i.e., the pH value of the doping solution) following the Nernst equation. As the pH decreases, the redox potential of BQ/HQ increases, making BQ a stronger p-dopant. This dependence of the redox potential of the BQ/HQ pair on pH is shown in Figure 1f in comparison to the oxidation potentials of different semiconducting SWNTs versus their diameter. Note that the redox potentials obtained for HiPco SWNTs from electrochemical photoluminescence quenching measurements[38] were extrapolated for the larger nanotube diameters, for which no experimental values are available. Based on these plots we may expect very limited p-doping for small-diameter SWNTs and efficient p-doping of large-diameter SWNTs, especially at low pH values.

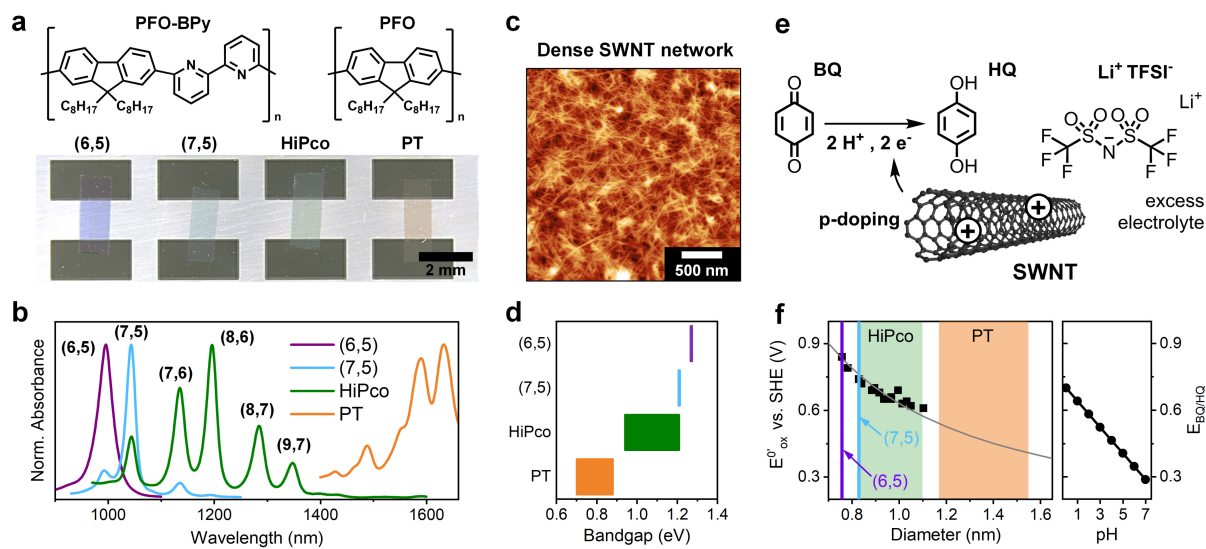

**Figure 1.** (a) Molecular structures of PFO-BPy and PFO as polymers for selective wrapping of semiconducting SWNTs and photograph of dense films of sorted (6,5), (7,5), HiPco, and PT-SWNT between gold electrodes on a glass substrate. (b) Absorption spectra showing the main $E_{11}$ peaks of the corresponding (6,5), (7,5), HiPco, and PT-SWNT dispersions. (c) Atomic force microscopy (AFM) image of a PT-SWNT network produced by vacuum filtration. (d) Bandgaps of (6,5) and (7,5) SWNTs and bandgap ranges of mixed diameter HiPco and PT-SWNTs. (d) Schematic of proton-coupled electron transfer doping with BQ as oxidant, SWNTs as semiconductor to be p-doped and TFSI as electrolyte anion for charge compensation. (f) Comparison of diameter-dependent redox potentials of different SWNTs (from ref. [38], left panel) and pH-dependent redox potential of BQ/HQ (right panel).

To test PCET doping of SWNTs, the different nanotube networks were immersed in aqueous and pH-buffered (pH 7 to 1) solutions of BQ (1 mM) with lithium bis(trifluoromethyl-sulfonyl)imide (Li[TFSI], 100 mM) as the excess electrolyte for 1 min. To ensure equal initial conditions before the first doping step, all samples were treated with ethylenediamine solution in ethanol to remove unintentional doping caused by ambient water/oxygen. The doping efficiency was monitored by near-infrared absorption spectroscopy of the films. A reduction of absorbance between undoped and doped films is a direct and even quantitative indicator for the charge carrier concentration (i.e. doping level) in semiconducting SWNTs.[26, 39] The characteristic bleaching of the main $E_{11}$ transition and the emergence of charge-induced red-shifted absorption peak (usually assigned as a trion, $T^+$) are clear signs of doping.[40]

When networks of (6,5) and (7,5) SWNTs were immersed consecutively in PCET doping solutions of decreasing pH, the $E_{11}$ absorption peak at 995 nm for (6,5) and at 1046 nm for (7,5) films was reduced and a trion peak emerged at approximately 1150 nm and 1210 nm, respectively (see Figure 2a,b). Although the oxidation potentials of (6,5) and (7,5) SWNTs are different, the values for $E_{11}$ bleaching only start to deviate at lower pH when the dopant is stronger (see Figure 2c). However, even at very low pH values the bleaching did not go beyond 35%, which indicates low p-doping levels of about 50 holes per µm.[26] This relatively low level of doping by PCET is also reflected in the limited electrical conductivities reached for (6,5) SWNT films even at pH 1 (~ 8.4 S cm$^{-1}$). Much higher $E_{11}$ bleaching (to nearly zero) and conductivities (up to few thousand S cm$^{-1}$) are usually obtained for (6,5) and (7,5) SWNT networks through doping with AuCl$_3$[26], OA[23, 41] or ion-exchange doping with AuCl$_3$/TFSI.[15]

The clear impact of the different bandgaps and oxidation potentials of semiconducting SWNTs can be demonstrated with the five different nanotubes species that are selected by polymer-wrapping from HiPco nanotubes (i.e., (7,5), (7,6), (8,6), (8,7) and (9,7) SWNTs) with bandgaps from 1.21 eV to 0.83 eV. For decreasing pH of the dopant solution, a clear difference in the bleaching levels for the five nanotube species is evident (see Figure 2d). When the absorption spectra are normalized to the $E_{11}$ peak (at 1052 nm) of the (7,5) nanotubes (largest bandgap), these differences become even more visible (see Figure 2e). The $E_{11}$ transitions of the (9,7), (8,7) and (8,6) nanotubes (at 1354, 1285, and 1205 nm, respectively) with the lowest oxidation potential are bleached much more than the (7,6) and (7,5) nanotubes especially at higher pH. The precise spectral analysis is complicated by the appearance of trion absorption peaks that overlap with the (9,7) and (8,7) nanotube transitions. Hence in Figure 2f, the reduction of the $E_{11}$ absorbances are only shown for the (7,5) SWNTs, (7,6) SWNTs (at 1134 nm), and (8,6)

SWNTs versus pH normalized to their initial undoped state. As expected, there is a clear correlation of the degree of doping by PCET with the pH of the doping solution and the oxidation potential of each nanotube species. The (8,6) SWNTs are doped 1.5 times as much as the (7,5) nanotubes for a difference in redox potential of ~80 mV.

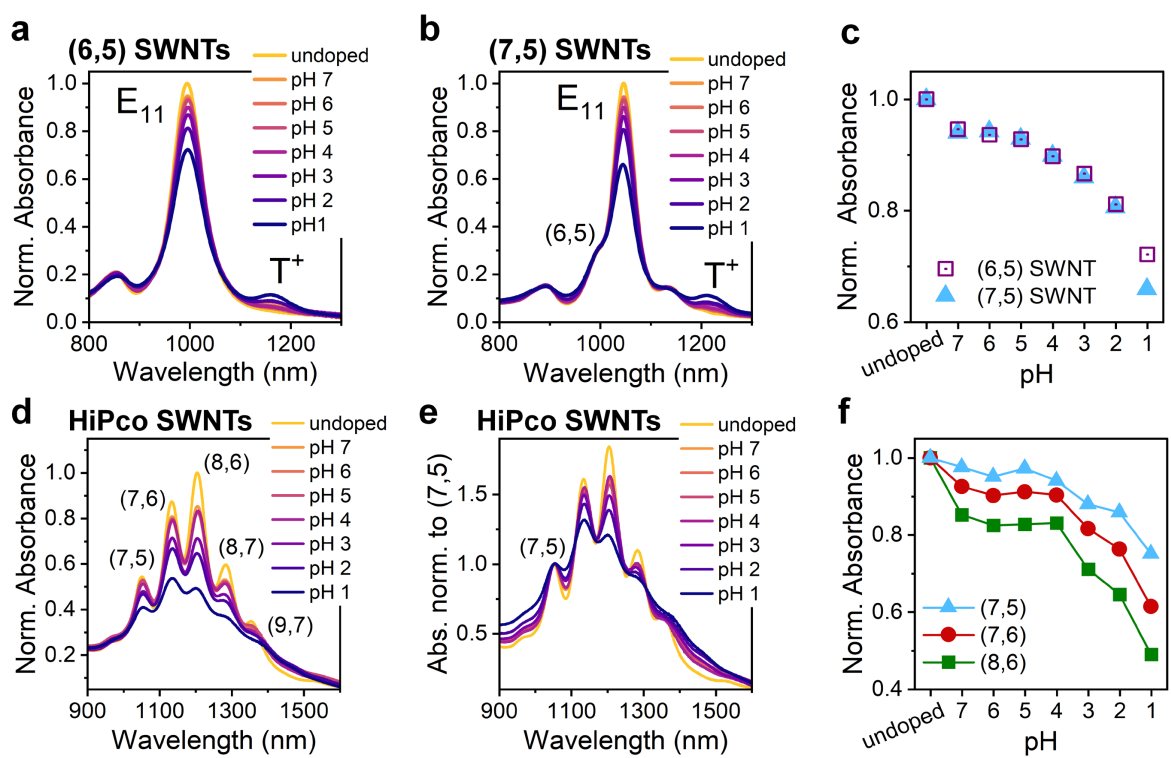

**Figure 2.** Absorption spectra of films of (a) (6,5) and (b) (7,5) SWNTs (here with residual (6,5) SWNTs) doped via PCET with BQ/TFSI at different pH values showing a progressive bleaching of the main $E_{11}$ absorption peaks. (c) Change of $E_{11}$ absorbance (normalized to undoped state) of (6,5) and (7,5) SWNT films depending on pH of doping solution. (d) Absorption spectra of a HiPco SWNT film PCET-doped at different pH values and (e) normalized to (7,5) SWNT absorption. (f) Change of $E_{11}$ absorbance (normalized to undoped state) of (7,5), (7,6) and (8,6) SWNTs in sorted HiPco nanotube film depending on pH value during PCET-doping.

This clear impact of the position of the valence band on doping of mixed small-diameter SWNTs was shown to a similar extent for electrochemically and electrostatically doped networks.[16, 40, 42] Such variations in doping levels and thus electrical conductivity (also determined by contact resistance between nanotubes)[43-44] in a network also cause the formation of specific percolation paths for charge carriers through smaller bandgap nanotubes,[45] which reduces the effective overall network density and hence conductivity. Indeed, the mixed HiPco network still only shows a maximum conductivity of 8.4 S cm$^{-1}$, although larger-diameter nanotubes exhibit higher carrier mobilities than small-diameter nanotubes.[7, 46]

While moderate p-doping levels can be achieved through PCET with BQ/TFSI for small-diameter nanotube networks, the obtained conductivity values are inferior to other chemical doping methods[15, 23, 26] and are not sufficient for thermoelectric applications.[2] However, based on Figure 1f, large-diameter PT-SWNTs with much smaller bandgaps and lower oxidation potentials should be doped efficiently via PCET with BQ. In terms of thermoelectric applications, the intrinsically higher carrier mobilities of large-diameter nanotubes,[7, 46] good Seebeck coefficients[3, 22] and higher chemical stability make them much more suitable materials than small-diameter nanotubes. This notion is confirmed by the absorption bleaching observed for dense films of polymer-sorted PT-SWNTs in Figure 3a. Upon PCET-doping at pH 1, the broad $E_{11}$ absorption around 1641 nm (originating from several large-diameter nanotubes) is strongly bleached to ~20% of its undoped value (corresponding to about 800 holes per micrometer).[39] Additionally, we observe a slight bleaching of the $E_{22}$ absorption around 940 nm as well as a rising broad absorbance band further in the near-infrared, which is consistent with previous reports on strongly doped SWNTs.[15, 47]

As the $E_{11}$ absorbance decreases nearly linearly with decreasing pH (see Figure 3b), the electrical conductivity of the PT-SWNT networks increases from ~5 S cm$^{-1}$ in the undoped state to a maximum of 360 S cm$^{-1}$ when doped at pH 1. While these values are an order of magnitude lower than those achieved by chemical doping of PT-SWNT networks, they are in the region where the maximum PF values for PT-SWNTs are observed.[3, 15] Interestingly, we find that mixtures of BQ and pH 1 buffer, without TFSI, do not dope the SWNTs (see Figure S2) although the buffer solution contains a significant concentration of phosphate anions. This may be due to their hydrophilicity and hydration shell, which prevents them from strongly interacting with the hydrophobic SWNTs and wrapping polymer. In contrast to that, the hydrophobic and larger TFSI anions can interact easily with the SWNTs, similar to what Ishii *et al*. observed for semiconducting polymers.[35]

An important difference to ion-exchange doping of nanotubes, which also typically employs TFSI as a counter ion, is the fact that doping by PCET occurs under ambient conditions, using water as a benign solvent and is reversible. The doping level can be changed several times through the application of doping solutions with different pH values to the same nanotube film. This is demonstrated for both the absorbance and the conductivity of a film of PT-SWNTs that was repeatedly doped at pH 1 and pH 3 (see Figure S3).

While the broad $E_{11}$ absorbance of PT-SWNTs in dense films does not allow for a simple analysis of preferential doping depending on oxidation potential, resonant Raman spectroscopy can give some insights. The intensity of many nanotube Raman modes is sensitive to charge carrier density,[48-49] either directly or through differences in resonant excitation.[50] We observe bandgap and excitation laser-dependent variations in the radial breathing mode (RBM) intensities of PCET-doped PT-SWNTs as shown and discussed in detail in the Supporting Information (Figure S4). Overall, we observe that the selection of the excitation wavelength controls which SWNT species are mainly probed by resonant Raman and that p-doping by PCET is similarly efficient for all PT-SWNTs. The latter is most likely due to the overall small bandgap and narrow bandgap range of PT-SWNTs (0.70–0.88 eV) compared to the HiPco nanotubes (0.94‑1.21 eV).

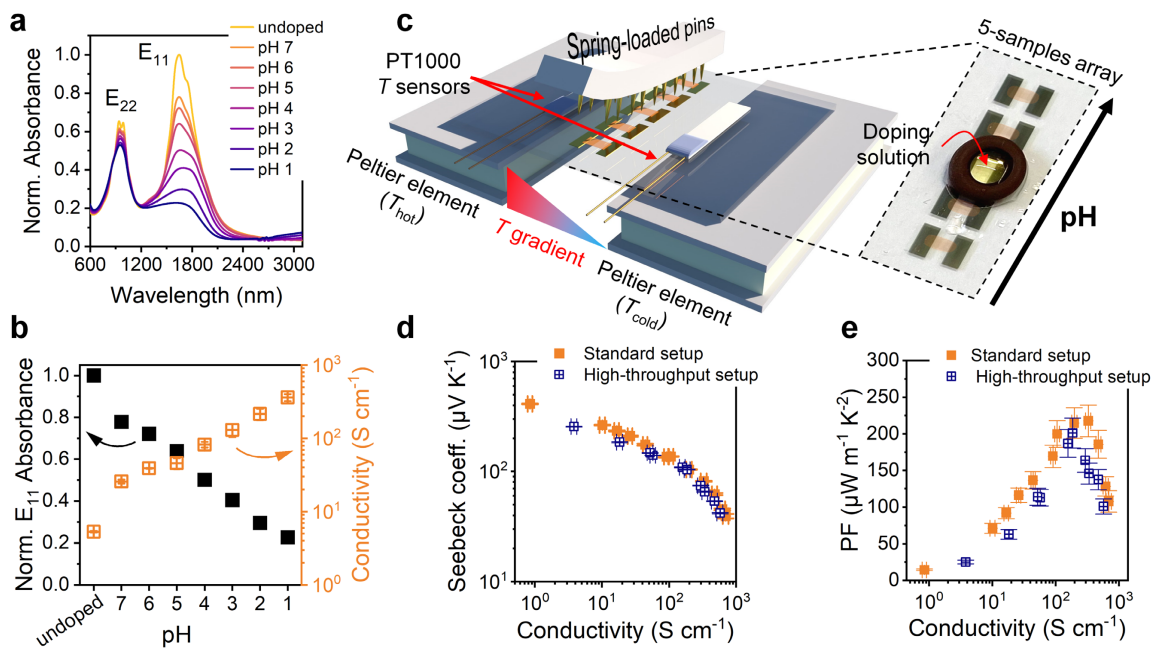

**Figure 3.** (a) Absorption spectra of PT-SWNT film doped via PCET with BQ/TFSI at different pH values. (b) Correlation between change of $E_{11}$ absorbance (normalized to undoped state) and conductivity for PCET-doped PT-SWNT film. (c) Schematic of a high-throughput setup for conductivity and Seebeck coefficient measurements and photograph of glass substrate with 5 pairs of gold electrodes and PT-SWNT films. The O-ring confines a pH-controlled, buffered dopant solution for local doping. (d,e) Seebeck coefficients versus conductivity and resulting PF values obtained from the standard measurement setup and the high-throughput setup for PT-SWNT networks doped by PCET.

To investigate the thermoelectric properties of SWNT networks after PCET doping at different pH values more efficiently and quickly, we developed a measurement setup to determine the electrical conductivities and Seebeck coefficients of up to five SWNT films on one substrate in parallel (schematically depicted in Figure 3c). The Seebeck coefficients of thin film semiconductors are usually measured with homebuilt setups employing either steady-state or quasi-steady state methods.[51-53] In both cases a temperature gradient ($\Delta T$) is created by heating/cooling one or both sides of the tested sample and the temperatures and thermovoltages ($\Delta V$) are measured to subsequently calculate the Seebeck coefficient as $S = \Delta V/\Delta T$. These measurements can be done either independently, using temperature sensors and sweeping current-voltage measurements[15] or the temperature and thermovoltages are measured simultaneously with thermocouples.[54] Samples are typically measured sequentially, which requires re-contacting of the electrodes resulting in long measurement cycles due to the time (3-5 min per temperature step) required to stabilize the temperature gradient in steady-state thermoelectric measurements. In contrast to that, the parallelized relay setup applied here uses the steady-state method to create a temperature gradient over up to five films between pairs of electrodes on a single glass substrate (Figure 3c). Each electrode pair is contacted by four spring pins (Kelvin geometry) connected to a relay board to switch the electrical connections. Further details are provided in the experimental section and the Supporting Information (Figure S5). Figure 3d,e shows the electrical conductivity, Seebeck coefficients and corresponding PF values for different PCET-doped PT-SWNT networks obtained with the relay setup in comparison to those from the standard setup (i.e., without relay). A maximum PF is reached with $218 \pm 21$ µW m$^{-1}$ K$^{-2}$ and $201 \pm 20$ µW m$^{-1}$ K$^{-2}$, respectively. The slight differences are within the instrumental error ranges and indicate that the relay setup is well-suited for accurate high-throughput measurements. Notably, these values are smaller than the largest power factors reported in the literature for polymer-wrapped SWNTs doped with dodecaborane clusters of 920 µW m$^{-1}$ K$^{-2}$,[3] which may be due to the specific dopant or the morphology of the SWNT films.

The relay setup can be used in two modes for high-throughput data acquisition. First, several different SWNT films can be treated with the same doping solution and fully characterized to eliminate undesired variations caused by different doping solutions. This is exemplified with thin films of (6,5), (7,5), HiPco, and PT-SWNTs on the same substrate after sequential PCET doping at different pH (see Figure S6), which show that the obtained PF values are mainly determined by the achievable doping levels for the different nanotube species. Second, films on the same substrate can be doped with different doping solutions and then characterized in parallel by using a simple O-ring to confine the doping solution (see Figure 3c). Aside from the improved comparability of measurements, the main advantage is the speed of data acquisition without the need for re-contacting devices. We estimate that typical thermoelectric measurements of an array of five different materials doped at six different doping levels performed with six temperature steps and 150 s as thermalization time would take an experienced researcher about 12.5 hours to complete (25 minutes per thermoelectric data point, assuming 5 minutes to dope, and 5 minutes to set up the measurement). Conversely, the same study would take only 2.5 hours (or 5 minutes per thermoelectric data point) with the relay setup, which represents an 80% reduction in experiment time over standard sequential measurements.

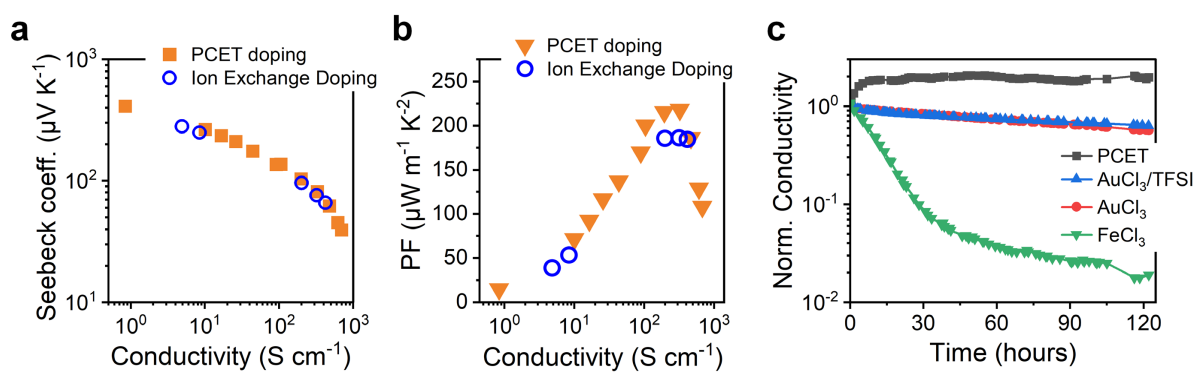

**Figure 4.** Seebeck coefficients versus conductivity (a) and resulting PF values (b) for both PCET-doped PT-SWNT films and those ion-exchange doped with $AuCl_3$/TFSI. (c) Comparison of long-term stability of the electrical conductivity of PT-SWNT films doped by PCET, $AuCl_3$, ion-exchange doped with $AuCl_3$/TFSI, and $FeCl_3$ (measured together in the high-throughput setup). All values are normalized to the initial conductivity after doping (PCET: 299 S cm$^{-1}$, $AuCl_3$/TFSI: 1131 S cm$^{-1}$, $AuCl_3$: 1298 S cm$^{-1}$, $FeCl_3$: 688 S cm$^{-1}$).

As shown above, PT-SWNT networks can be efficiently doped by PCET with BQ/TFSI, especially at low pH values. The observed electrical conductivities, Seebeck coefficients and resulting PFs are indeed equal to those achieved by ion-exchange doping with $AuCl_3$/TFSI as shown in Figure 4b,c. In both cases a clear maximum of the PF values is observed and stronger doping would actually be detrimental to the thermoelectric properties. Clearly, PCET has the practical advantage of being performed with aqueous solutions in air rather than in organic and anhydrous solvents in a glovebox. It also provides very reproducible doping levels, which can be adjusted easily by the pH value of a buffered electrolyte solution.

However, an even more important aspect is the long-term stability of the conductivity of the doped films. Only very few studies discuss the stability of p-doping in SWNT networks[15, 27] or even organic semiconductors,[34, 55] although it is recognized as a common problem. PT-SWNT networks exhibit an almost two orders of magnitude decrease in conductivity over a period of 120 hours when strongly doped with $FeCl_3$ as shown in Figure 4c. Raman measurements (Figure S7) reveal a significant increase in the $D/G^+$ ratio for $FeCl_3$-doped SWNTs, indicating an increased number of lattice defects,[56] that hinder charge transport[14]. Radicals formed from the dopant counterions presumably attack the $sp^2$ carbon lattice, as demonstrated by Jha *et al.* for polymer semiconductors doped by $FeCl_3$.[29]

$AuCl_3$ doping and ion-exchange doping with $AuCl_3$/TFSI also lead to a noticeable reduction of conductivity over five days and a slight increase in the Raman $D/G^+$ ratio. In stark contrast to that, the conductivity of PCET-doped PT-SWNTs nearly doubled within the first 10 hours after doping and then remained stable over the entire measurement period, with only minor fluctuation. The initial increase in conductivity might be explained with achieving more uniform doping throughout the entire network including nanotube bundles over time or continued doping by residual BQ in the dense film. The excellent stability of the doped PT-SWNT films over the subsequent measurement period may be due to the properties of HQ as a radical scavenger.[57] Irrespective of the precise mechanism, this excellent long-term stability of PCET-doped films of PT-SWNTs, combined with the facile application and tunability of PCET to reach maximum PFs makes this doping method highly suitable for future thermoelectrics based on carbon nanotubes.

## 3. Conclusions

We demonstrated PCET-doping of dense films of different polymer-sorted semiconducting SWNTs with BQ/TFSI and found a clear dependence of the doping efficiency on the bandgap and oxidation potential of the nanotubes. Near-infrared absorption spectra reveal the carrier concentration in doped SWNTs films, which can be controlled by the proton concentration in the doping solution (i.e. the pH value). As the bandgap of the SWNTs decreases with increasing diameter, BQ changes from being a mild dopant for (6,5) and (7,5) SWNTs to an efficient dopant for PT-SWNTs. A fast comparison of the thermoelectric properties of up to five different PCET-doped SWNT films at different pH values was enabled by a relay measurement setup that reduced the experiment time by 80 %. The main limiting factor for the PF values of the different nanotube films proved to be the achievable doping level and hence conductivity. Only small bandgap PT-SWNTs were doped sufficiently at low pH to reach maximum PF values that were equal to those obtained by ion-exchange doping with $AuCl_3$/TFSI. PCET doping offers several advantages over doping with strong oxidants like $AuCl_3$, $FeCl_3$, and even ion-exchange doping. It can be performed reversibly with aqueous solutions in air and - most importantly - PCET-doped nanotube networks exhibit excellent long-term stability. Overall, PCET doping represents an important advancement for p-doping of semiconducting SWNT networks as resilient and high-performing materials for thermoelectrics.

## 4. Experimental Section

**Materials**

CoMoCAT SWNTs (CHASM Advanced Materials Inc. SG65i), HiPco SWNTs (Unidym Inc.), PT-SWNTs (Raymor Industries Inc., RN-220), poly(9,9-dioctylfluorenyl-2,7-diyl) (PFO, Sigma Aldrich, $M_w \geq 20$ kg mol$^{-1}$), poly[(9,9-di-*n*-octylfluorene-2,7-diyl)-*alt*-(2,2′-bipyridine-6,6′-diyl)] (PFO-BPy, American Dye Source, $M_w \approx 40$ kg mol$^{-1}$) were used for the selective dispersion of semiconducting SWNTs. The dopants 1,4-benzoquinone (BQ) (≥98.0%) and hydroquinone (HQ) (≥99.0%) were purchased from Tokyo Chemical Industry (TCI). Dopants iron(III) chloride (FeCl$_3$) (≥99.99%) and gold(III) chloride (AuCl$_3$) (≥99.99%) and de-dopant ethylenediamine were purchased from Sigma-Aldrich. FeCl$_3$ doping and ion-exchange doping with AuCl$_3$:[BMP][TFSI] and were performed using anhydrous acetonitrile (<10 ppm water) as solvent. The counterion salts lithium bis(trifluoromethylsulfonyl)imide (Li[TFSI]) (99.95%) and 1-butyl-1-methylpyrrolidinium bis(trifluoromethylsulfonyl)imide ([BMP][TFSI]) (≥98.5%, <0.0400% water) were purchased from Sigma-Aldrich. Phosphoric acid (H$_3$PO$_4$), potassium dihydrogen phosphate (KH$_2$PO$_4$), acetic acid (CH$_3$COOH), potassium acetate (CH$_3$COOK), ammonium sodium phosphate dibasic tetrahydrate (NaNH$_4$HPO$_4 \cdot 4$ H$_2$O) were used as received from Sigma-Aldrich to make buffer stock solutions. Potassium hydroxide (KOH) and sulfuric acid (H$_2$SO$_4$) were added to adjust the pH of the buffer stock solutions.

*Selective Dispersion of semiconducting SWNTs*

Dispersions of (6,5) and (7,5) SWNTs were prepared by adding 0.4 g L$^{-1}$ or 0.5 g L$^{-1}$ of CoMoCAT SWNTs, respectively, to a 0.5 g L$^{-1}$ PFO-BPy (for (6,5) SWNTs) or 0.9 g L$^{-1}$ PFO solution (for (7,5) SWNTs) in toluene, followed by shear-force mixing for 72 h (Silverson L2/Air, 10230 rpm, 20 °C). Sorted PT-SWNT dispersions were obtained by shear-force mixing 1.5 g L$^{-1}$ of the RN-220 raw material in a solution of 0.5 g L$^{-1}$ PFO-BPy in toluene under the same conditions. Details on the shear-force mixing process can be found elsewhere.[11, 58] Dispersions of HiPco SWNTs were prepared by bath sonicating 1.5 g L$^{-1}$ HiPco raw material and 2.0 g L$^{-1}$ PFO in toluene for 45 mins, as described previously.[40] All dispersions were centrifuged twice for 45 min at 60,000 g (Beckman Coulter Avanti J26SXP centrifuge) and the supernatant was collected. The supernatants of the (6,5), (7,5), and HiPco SWNTs were filtered through a poly(tetrafluoroethylene) (PTFE) syringe filter (Whatman, pore size 5 μm) to eliminate large aggregates. Excess wrapping polymer was removed by vacuum filtration of the dispersions through a PTFE membrane (Merck Omnipore, JVWP, pore size 0.1 μm, diameter 25 mm). The obtained SWNT filter cake was subsequently soaked in 80 °C fresh toluene (five times, 5 min each) before being redispersed in pure toluene by bath sonication for 30 min. At this point, the dispersions were characterized by UV–Vis–NIR absorption spectroscopy.

*SWNT Film Deposition and Characterization*

To obtain dense and uniform SWNT films, the diluted dispersions were first passed through a PTFE syringe filter (Whatman, pore size 5 μm) and then vacuum filtered (approx. 20 mL) through an MCE membrane filter (Merck MF-Millipore, VSWP, pore size 0.025 μm, diameter 25 mm). Circular films were cut from the MCE membrane using a rotary punch, while rectangular films were cut with a razor blade. The SWNT side of the MCE membrane was positioned on a substrate, wet with 2-propanol, covered with aluminum foil, and manually pressed onto the substrate. The MCE membrane was subsequently dissolved in acetone. For optical measurements, SWNTs were deposited on quartz substrates (UQG Optics, FQP-1212). For electrical measurements, glass substrates (Schott AF32eco, 20 × 25 × 0.5 mm) with pairs of thermally evaporated Cr (2 nm)/Au (20 nm) electrodes with 1.5 mm spacing were used. To account for the sample geometry when calculating the electrical conductivity, film lengths and widths were measured with an Olympus BX51 microscope and the thickness was determined with a Bruker DektakXT Stylus profilometer. Atomic force microscopy images were recorded with a Bruker Dimension Icon atomic force microscope in ScanAsyst mode.

**Doping procedure**

Aqueous PCET doping solutions were prepared by mixing stock solutions of pH buffer, Li[TFSI], and BQ immediately before doping to a concentration of $(10:100:1) \times 10^{-3}$ M of pH buffer:Li[TFSI]:BQ. To dope past the maximum PF in Figure 3d,e, solutions with a higher concentration of electrolyte and BQ, i.e., $(10:1000:50) \times 10^{-3}$ M of pH buffer:Li[TFSI]:BQ at pH 1 – 4 were applied. For reversible doping in Figure S3, solutions containing HQ were used at a concentration of $(10:100:10:10) \times 10^{-3}$ M of pH buffer:Li[TFSI]:BQ:HQ. Doping was performed by covering films with 200 μL of doping solution for 1 minute, removing the solution with a pipette and drying the films with nitrogen. Electrical and optical measurements were performed immediately after doping each sample. To maintain consistency within each measurement series, doping was performed on SWNT films cut from the same filter membrane and using the same stock solutions of pH buffer, BQ, and electrolyte. AuCl$_3$, FeCl$_3$ and ion-exchange doping (with AuCl$_3$/[BMP][TFSI]) were performed inside a dry nitrogen glovebox. The dopants were dissolved in acetonitrile. SWCNT films between prepatterned gold electrodes on glass were covered with 250 μL of the dopant solution for 5 minutes on a spin coater. The films were dried by spinning at 8000 rpm and washed with 250 μL of acetonitrile while spinning, as described previously.[15] Before performing UV-Vis-NIR absorption measurements, all SWNT films were immersed in a 1% (v/v) ethylenediamine solution in ethanol for 1 min to remove unintentional p-doping caused by ambient water/oxygen. Subsequently, the films were thoroughly washed with ethanol to remove the ethylenediamine.

*Spectroscopic Characterization*

UV–Vis–NIR absorption spectra of SWNT films on quartz substrates were collected with a Varian Cary 6000i or Agilent Cary 5000 spectrometer. Raman spectra were collected with a Renishaw inVia Reflex confocal Raman microscope in the backscattering configuration equipped with a 50x objective (N.A. 0.5). For each spectrum, >200 individual spectra from an area of 100 μm$^2$ were collected and averaged. Excitation wavelengths: 532, 633 and 785 nm.

*Thermoelectric Characterization*

Seebeck coefficients ($S$) were measured with a previously described homebuilt setup.[15] A glass substrate with the SWNT film positioned between a pair of electrodes was suspended between two Peltier elements and the surface temperature of the glass substrate was measured at each side with PT1000 temperature sensors. For six temperature differences ($\Delta T$) within ±1 K across the SWNT film, current-voltage sweeps (±10 mV) were recorded with a Keysight B1500A semiconductor parameter analyzer and used to extract thermovoltages $\Delta V$. The Seebeck coefficient ($S = \Delta V/\Delta T$) was calculated from a linear fit of the thermovoltages versus the temperature difference. The electrical conductivity ($\sigma$) was calculated for each of the six current-voltage sweeps and averaged. The power factor was calculated as PF = $S^2\sigma$.

This setup was adapted for high-throughput measurements on substrates with 5 pairs of electrodes and 5 SWNT films in parallel. The electrodes were contacted with 20 spring pins (see Figure 3c). The force and sense high and low connections of the Keysight B1500A used for the current-voltage sweeps were connected to a relay board with 16 mechanical relays (Seeit TTL-RELAY16-5V) controlled by an Arduino Mega 2560 Rev 3 programmed with a custom C++ script to accept commands to switch the relays from a main Python script for Seebeck coefficient measurements. At each temperature difference, the current-voltage measurements were performed consecutively for the five devices and the temperature difference before and after the current-voltage sweep for each device was recorded and averaged.

**Supporting Information**

Supporting Information is available from the Wiley Online Library or from the author.

**Acknowledgements**

This project has received funding from the European Union's Horizon 2020 research and innovation program under the Marie Skłodowska-Curie grant agreement No. 955837 (HORATES). X.R.M acknowledges funding from the Alexander von Humboldt Foundation. The authors thank Klaus Schmitt and Günter Meinusch for their support in building the high-throughput measurement setup. The authors also thank Zia Ullah Kahn and Penghui Ding for their help in the first experiments for this project.

Received: ((will be filled in by the editorial staff))
Revised: ((will be filled in by the editorial staff))
Published online: ((will be filled in by the editorial staff))

# Supporting Information

# Bandgap-Dependent Doping of Semiconducting Carbon Nanotube Networks by Proton-Coupled Electron Transfer for Stable Thermoelectrics

*Angus Hawkey, Xabier Rodríguez-Martínez\*, Sebastian Lindenthal, Moritz C. F. Jansen, Reverant Crispin, Jana Zaumseil\**

Angus Hawkey, Xabier Rodríguez-Martínez, Sebastian Lindenthal, Moritz C. F. Jansen, Jana Zaumseil

Institute for Physical Chemistry, Heidelberg University, 69120 Heidelberg, Germany

Xabier Rodríguez-Martínez

Universidade da Coruña, Centro de Investigación en Tecnoloxías Navais e Industriais (CITENI), Ferrol, 15403, Spain

Reverant Crispin

Laboratory of Organic Electronics, Department of Science and Technology, Linköping University, Sweden

E-mail: zaumseil@uni-heidelberg.de; xabier.rodriguez@udc.es

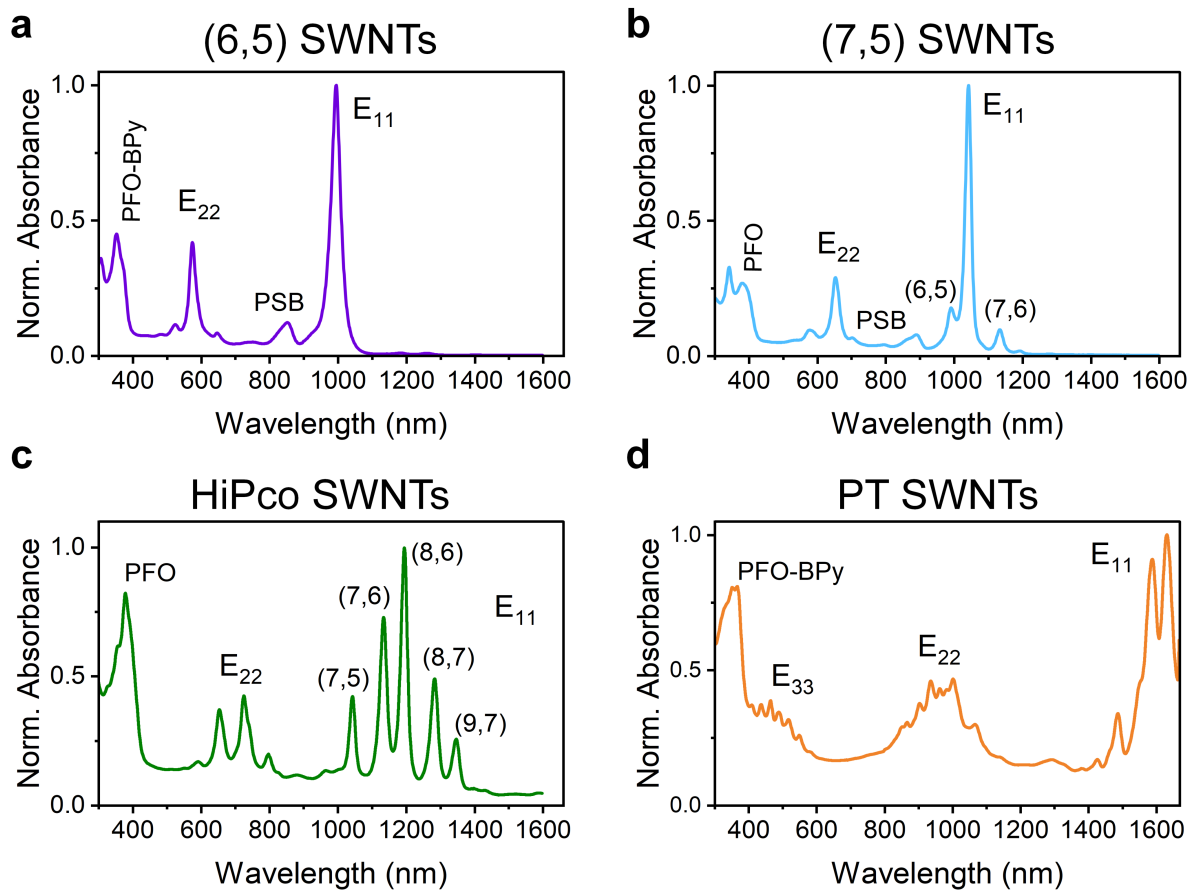

**Figure S1.** Normalized absorption spectra of polymer-sorted dispersions of (a) (6,5), (b) (7,5), (c) HiPco, and plasma torch (PT) SWNTs in toluene.

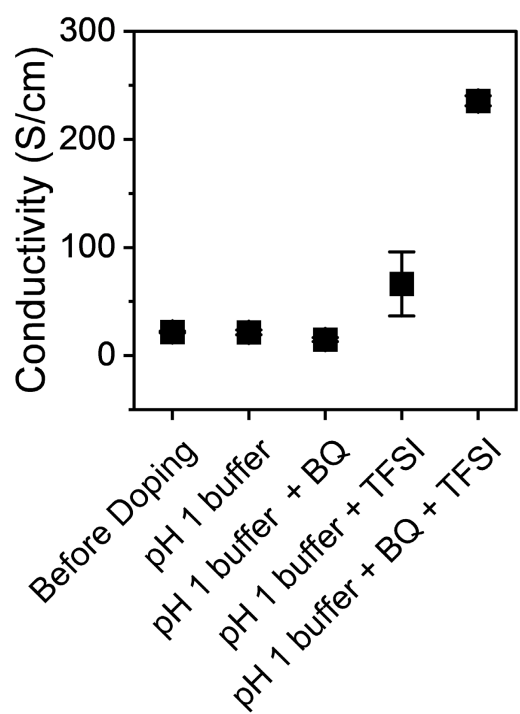

**Figure S2.** Electrical conductivity of PT SWNT films after immersion in the different solutions for 1 minute and drying with compressed nitrogen. The pH buffer is 10 mM of $H_3PO_4/KH_2PO_4$. BQ and TFSI were used at 50 mM and 100 mM, respectively.

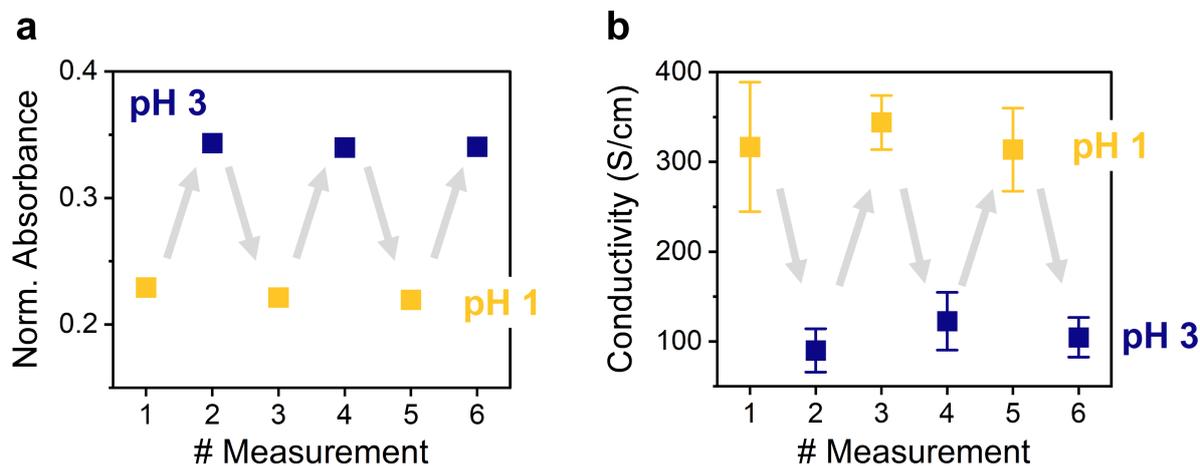

**Figure S3.** Demonstration of reversibility of PCET doping of a PT-SWNT film. The SWNT film was repeatably immersed in BQ/HQ/TFSI solutions of pH = 1 and then pH = 3. The maximum absorbance at $E_{11}$ (normalized to undoped state) (a) and the conductivity of the film (b) changed accordingly and reproducibly several times.

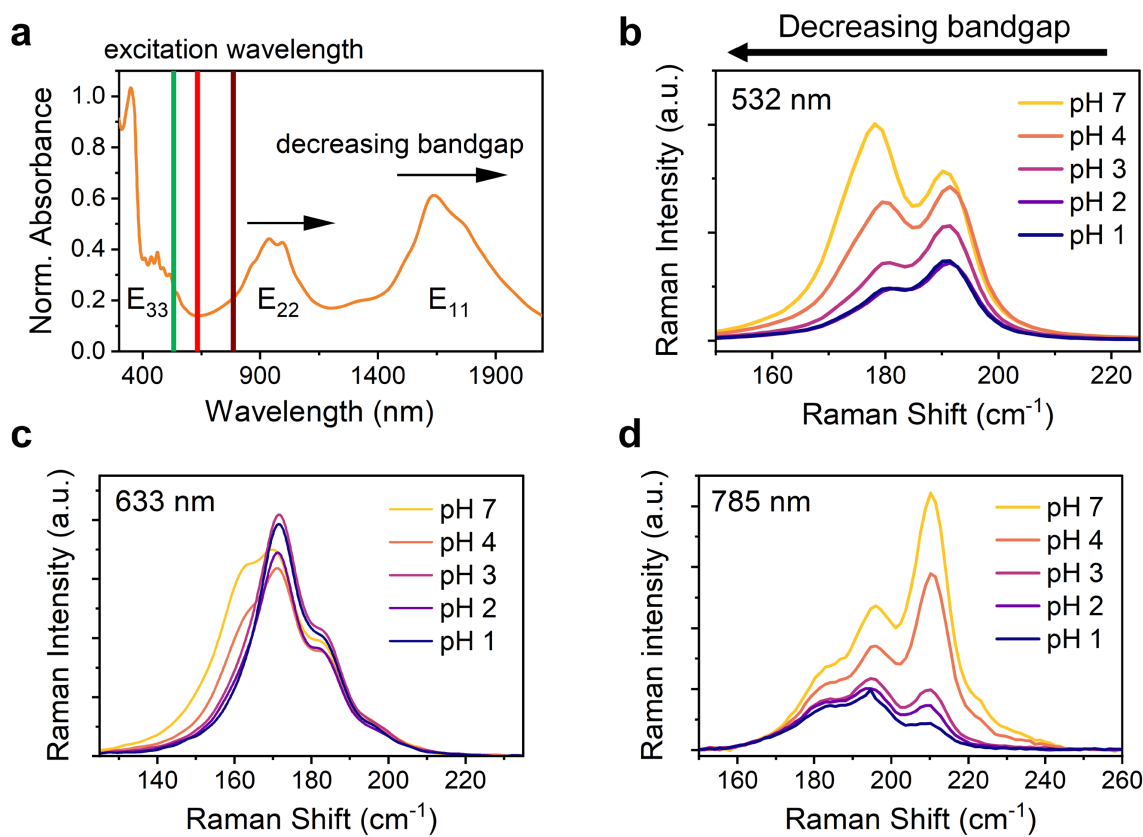

**Figure S4.** Doping of the PT-SWNTs can be characterized by resonant Raman spectroscopy. The wavenumber of the radial breathing modes (RBM) is inversely proportional to the SWNT diameter and hence correlates directly with the nanotube bandgap. The intensities of the RBM peaks depend on the overlap of the excitation laser wavelength with the nanotube's absorption bands (resonant excitation). Thus, they decrease with increasing doping as absorption is bleached. (a) Absorption spectrum of a network of sorted PT-SWNTs. The wavelengths of the three Raman lasers either overlap with the $E_{33}$ absorption of the smaller bandgap nanotubes (532 nm), only with the residual metallic nanotubes (633 nm) or with the $E_{22}$ absorption of the larger bandgap nanotubes (785 nm). (b) The smaller bandgap chiralities are more resonant with the 532 nm laser meaning the intensities of the RBM peaks at lower wavenumbers (~180 cm$^{-1}$) decrease more significantly with doping (here lower pH) than those at higher wavenumbers (~190 cm$^{-1}$). (c) The RBM peaks of metallic nanotubes and off-resonance excited semiconducting SWNTs barely change in intensity. (d) When the nanotubes of the high-energy flank of the $E_{22}$ absorbance are resonantly excited and this absorbance is bleached through doping, the RBM intensities of the large-bandgap SWNTs (~210 cm$^{-1}$) change the most as they absorb closer to the 785 nm laser (i.e. they are more resonant).

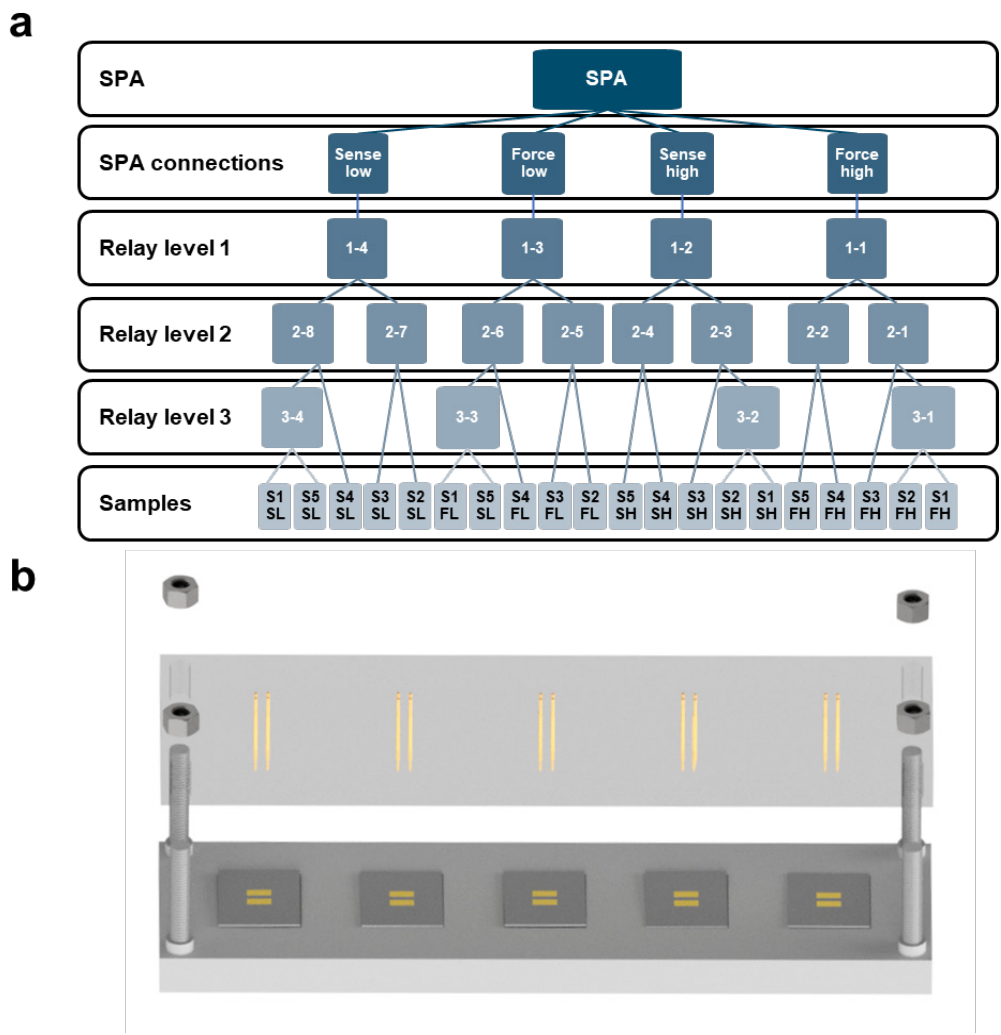

**Figure S5.** (a) Schematic diagram of the relay board used to switch connection between the semiconductor parameter analyzer (SPA) and five pairs of electrodes. This relay board is used for both the high-throughput electrical conductivity and Seebeck coefficient measurement setup, as well as the stability monitoring setup shown in (b), which tracks only the electrical conductivity over time. SL, FL, SH, and FH represent the connection of the samples to the sense low, force low, sense high, and force high lines, respectively, of the SPA through the relay board. (b) stability monitoring setup for tracking 4-probe electrical conductivity over time. Four spring contacts connect each of the five electrode pairs to a source meter using the relay setup in (a).

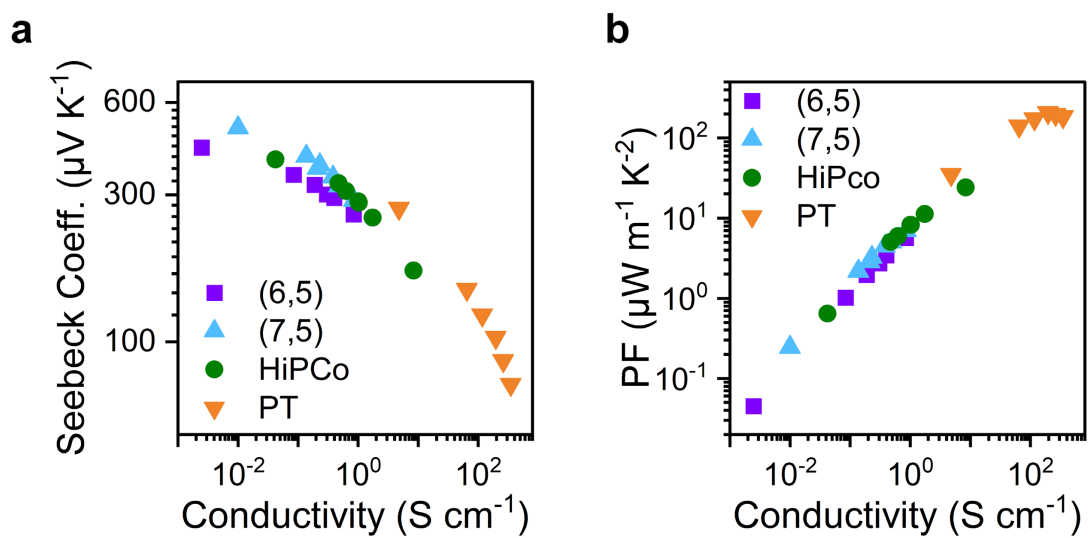

**Figure S6.** Electrical conductivities, Seebeck coefficients (a) and power factors (PF) (b) for different sorted semiconducting nanotube films obtained with the high-throughput measuring setup (here (6,5), (7,5) HiPco and PT- SWNTs) from the same substrate under the same PCET-doping conditions. The values for the PFs are mainly limited by the achievable doping levels for PCET doping. The highest PF values (207 $\mu W\ m^{-1}\ K^{-2}$) are obtained for the small-bandgap PT-SWNTs.

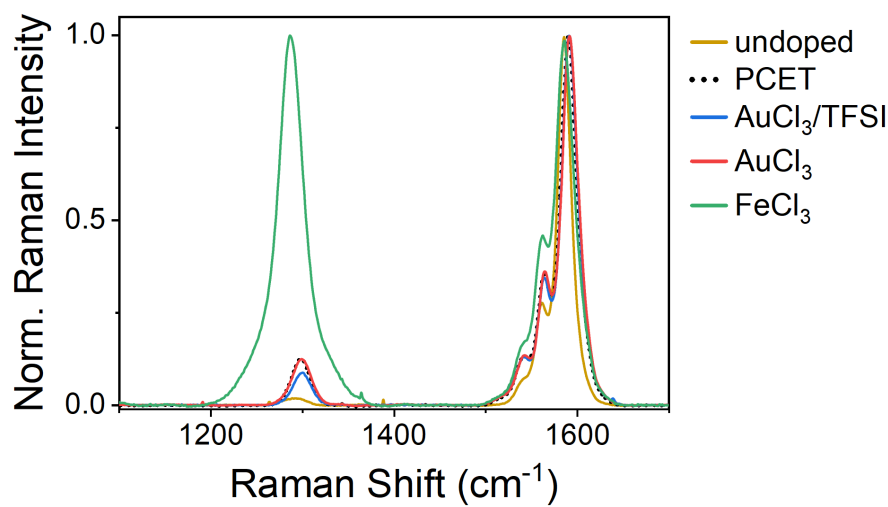

**Figure S7**. Raman measurements of PT-SWNTs (excitation at 785 nm) 120 hours after doping with $FeCl_3$, $AuCl_3$, ion-exchange doped ($AuCl_3$/TFSI), and PCET doped (corresponding to samples in Figure 4c). A large increase of the D-peak intensity relative to the $G^+$ peak in the $FeCl_3$-doped SWNTs indicates a drastic rise in the number of lattice defects in the PT-SWNTs leading to reduced carrier mobilities. For the other doping methods, the introduction of lattice defects is limited and the reduction in conductivity seems to arise from partial de-doping.